\documentclass[twocolumn,prb,aps,showpacs,superscriptaddress]{revtex4-1}
\usepackage{graphicx}
\usepackage[dvipdfm]{hyperref}

\newcommand{\beq}{\begin{equation}}
\newcommand{\eeq}{\end{equation}}

\begin{document}
\title{Magnetic strong coupling in a spin-photon system and transition to classical regime}

\author{I. Chiorescu}
\affiliation{Department of Physics and The National High Magnetic Field Laboratory, Florida State University, Tallahassee, Florida 32310, USA}

\author{N. Groll}
\affiliation{Department of Physics and The National High Magnetic Field Laboratory, Florida State University, Tallahassee, Florida 32310, USA}

\author{S. Bertaina}
\affiliation{Department of Physics and The National High Magnetic Field Laboratory, Florida State University, Tallahassee, Florida 32310, USA}
\affiliation{IM2NP-CNRS (UMR 6242), Universit\'{e} Aix-Marseille, 13397 Marseille Cedex, France.}

\author{T. Mori}
\affiliation{Department of Physics, Graduate School of Science, The University of Tokyo, 7-3-1 Hongo, Bunkyo-ku, Tokyo 113-8656, Japan}
\affiliation{CRESTO, JST, 4-1-8 Honcho Kawaguchi, Saitama 332-0012, Japan}

\author{S. Miyashita}
\affiliation{Department of Physics, Graduate School of Science, The University of Tokyo, 7-3-1 Hongo, Bunkyo-ku, Tokyo 113-8656, Japan}
\affiliation{CRESTO, JST, 4-1-8 Honcho Kawaguchi, Saitama 332-0012, Japan}

\date{Received 8 April 2010, published 14 July 2010, Phys Rev B {\bf 82}, 024413 (2010)}%

\begin{abstract}
We study the energy level structure of the Tavis-Cumming model applied to an ensemble of independent magnetic spins $s=1/2$ coupled to a variable number of photons. Rabi splittings are calculated and their distribution is analyzed as a function of photon number $n_{\rm max}$ and spin system size $N$. A sharp transition in the distribution of the Rabi frequency is found at $n_{\rm max}\approx N$. The width of the Rabi frequency spectrum diverges as $\sqrt{N}$ at this point. For increased number of photons $n_{\rm max}>N$, the Rabi frequencies converge to a value proportional to $\sqrt{n_{\rm max}}$. This behavior is interpreted as analogous to the classical spin-resonance mechanism where the photon is treated as a classical field and one resonance peak is expected. We also present experimental data demonstrating cooperative, magnetic strong coupling between a spin system and photons, measured at room temperature. This points towards quantum computing implementation with magnetic spins, using cavity quantum-electrodynamics techniques. 
\end{abstract}


\maketitle

\section{introduction}

Interactions of quantum systems with electromagnetic excitations are at the core of quantum information processing. Using photons and photonic entanglement, qubits can be detected and manipulated, and quantum information can, in principle, be transferred over long distances~\cite{Cirac_PRL97}. Of particular interest are the resonant modes in electromagnetic cavities, which have the potential of inducing a strong coupling regime such that the interaction outlast both photon's decay and qubit decoherence times~\cite{HarochePRL83,KimblePRL98}. Following the work of Dicke~\cite{DickePR54} on multi-atom superradiance, the case of a single atom in interaction with $n$ photons has been studied theoretically by Jaynes and Cummings~\cite{JCIEEE63}, and later on generalized \cite{TCPR1968,Cummings_PRA83,Agarwal_PRL84,Bogoliubov_JPA96}  for a number of $N$ otherwise non-interacting atomic systems. Other theoretical studies, applied to solid state systems~\cite{Trif_PRL2008}, have included environmental effects as well (e.g., in semiconducting materials~\cite{Houdre_PRA96,Piermarocchi_PRB08}) or ensemble-locking in a giant spin~\cite{Flatte_PRL10}. Experimentally, the phenomena of strong coupling regime has been reached by using the electric field component of the electromagnetic excitations: in one or more atomic systems \cite{Herskind_nphys09,Aoki_nat06}, semiconductors \cite{Khitrova_nphys06} and superconducting qubits in interaction with one~\cite{Wallraff_nat04} or more photons~\cite{Fink_nat08,Bishop_nphys09}. These studies prove the appearance of the so-called vacuum-field Rabi splitting (VRS) in the absorbtion peak of a probing photon field.

In contrast, achieving large magnetic coupling between a photon and a quantum spin, has been explored to a lesser extent, due to the typical smallness of the magnetic component (B-field) of the electro-magnetic field. However, since spin-based qubits do reveal significant coherence times for temperatures up to ambient value \cite{Nellutla_prl07, Bertaina_prl09}, the issue of coupling spin qubits to photons for data manipulation and transfer becomes of increasing interest. In the usual magnetic resonance methods, {\it e.g.} electron spin resonance (ESR), the absorbtion measurement of the electromagnetic field is related to the energy structure of the spin system. Feedback effects of the B-field component on the spins are ignored which means that for a two-level system there is one absorbtion peak at a frequency matching the levels separation. On the other hand, as mentioned above in the case of electrical coupling, photon absorption can probe the quantum mechanical interaction between the quantum system and the cavity photons which leads to VRS. In this paper, we will study the relation between these two cases, one classical and the other one quantum, by comparing the effects of magnetic coupling between $N$ non-interacting spins $s=1/2$ and the external radiation field. The transition between the classical and quantum case will be analyzed as well.

Because the wave length of the external field is large compared with
the distances between spins, all the spins interact with a 
single mode of electromagnetic field. In the classical case of spin resonance, the system Hamiltonian is given by ${\cal H}_{\rm S}=\cal H_{\rm Z0}+\cal H_{\rm ac}$ where $\cal H_{\rm Z0}$ is the Zeeman coupling to a static field $H_z$:
\begin{equation}
{\cal H}_{\rm Z0}=-\mu_0H_z\sum_{i=1}^Nm^z_i ={\hbar\omega_0\over2}\sum_i^N\sigma^z_i
\end{equation}
where $m_z=-g_s\mu_B\sigma^z_i/2$ gives the magnetic moment of spin $i$ ($g_s$ is the g-factor, equal to 2 for a free spin and $\mu_B$ is the Bohr magneton) and $\hbar\omega_0$ is the Zeeman splitting generated by $H_z$. The term $\cal H_{\rm ac}$ represents the spin coupling to an alternating B-field component $h_0$ oscillating with a frequency $\omega/2\pi$. Using the notation $\hbar\Omega_R=\mu_0h_0g_s\mu_B/2$ and the Pauli projection and raising/lowering operators this term is written as:
\begin{equation}
{\cal H}_{\rm acR} = 
{1\over2}\hbar\Omega_R\sum_{i=1}^N \left( e^{i\omega t}\sigma_i^+ + e^{-i\omega t}\sigma_i^-\right),
\label{ESR-RF}
\end{equation}
for a rotating B-field or, for an uniaxial B-field, as
\begin{equation}
{\cal H}_{\rm acX} =\hbar\Omega_R\cos(\omega t)\sum_{i=1}^N \sigma_i^x.
\label{ESR-MX}
\end{equation}

When treating the radiation field quantum mechanically, as in the Jaynes-Cummings model, the spin-photon coupling is described by a parameter $g$, here assumed to be the same for all spins:
\begin{equation}
{\cal H} =H_{\rm Z0}
+\hbar g\sum_i^N \left(b\sigma_i^+ + b^{\dagger}\sigma_i^-\right)
+\hbar\omega b^{\dagger}b,
\label{JC-R}
\end{equation}
with $b,b^{\dagger}$ the photon annihilation/creation operators. The amplitude of the electro-magnetic field is a dynamical variable, but not an external parameter as in the case of Eq. (\ref{ESR-RF}) or (\ref{ESR-MX}).
Namely, the strength of the electromagnetic field is given by $b^{\dagger}b$
in the quantum mechanical model in Eq. (\ref{JC-R}), while it is given by $h_{0}$ in the classical models in Eqs. (\ref{ESR-RF}) and (\ref{ESR-MX}).

\section{Complex susceptibility in the linear response theorem}

In the linear response theory~\cite{Kubo_JPSJ57,Abragam}, the imaginary part of the complex susceptibility is given by
\begin{equation}
\chi ''(\omega)={1-e^{-\beta\omega\hbar}\over 2}\int_{-\infty}^{\infty}
\langle M_x(0)M_x(t)\rangle_0 e^{-i\omega t}dt,
\label{chi}
\end{equation}
where $M_x={1\over2}\sum_i\sigma_i^x$, $\langle \cdots\rangle_0=
{{\rm Tr}\cdots e^{-\beta {\cal H_{\rm Z0}}}/{\rm Tr}e^{-\beta {\cal H_{\rm Z0}}}}$, $\beta=1/k_{\rm B}T$, $k_{\rm B}$ is the Boltzmann constant and $T$ is system's temperature.
This gives a coefficient of proportionality between the induced quantity 
$\langle M_x\rangle$ and the field $h_{0}$ at $h_{0}=0$. The eigenvalues of $\cal H_{\rm Z0}$ are simply given by $E_{k}=k\hbar\omega_0$, $k=-N,\cdots, N$, each of which is $_NC_k=N!/k!(N-k)!$ times degenerate. There is only one energy difference $\Delta E=\hbar\omega_0$ that has nonzero matrix element of $M_x$. In this case, each spin interacts only with the field, individually. Thus, we have a single peak in $\chi ''(\omega)$ at $\omega=\omega_0$. If we include some interactions among spins, such as the dipole-dipole interactions, the degeneracy of the energy levels of the $N$ spin system would be resolved, and additional peaks in $\chi"(\omega)$ are expected. 

\subsection{Quantum dynamics of paramagnetic spins under an ac field: Rabi oscillation}

If we consider the dynamics of spins in the ESR Hamiltonian ${\cal H}_{\rm S}$, 
the total magnetization shows the so-called Rabi oscillation. 
By transforming the wave function $|\Phi(t)\rangle$ as 
\begin{equation}
|\Psi(t)\rangle=e^{i{1\over2}\omega_0t\sigma_z}|\Phi(t)\rangle\equiv U|\Phi(t)\rangle,
\end{equation}
the rotating frame version of ${\cal H}_{\rm S}$ is given by 
\begin{equation}
U{\cal H}_{\rm S}U^{-1} = {1\over2}\hbar\Omega_R\sum_i^N \left(\sigma_i^+ 
+ \sigma_i^-\right)=\hbar\Omega_R\sum_i^N\sigma_x,
\label{ESR-RTF}
\end{equation}
at resonance. In this representation, the $z$ component of the magnetization rotates around the $x$-axis with the angular velocity ${\dot\theta=\Omega_R}$, which is the Rabi oscillation. It should be noted that the phase oscillates with the angular frequency $\omega_0$, which causes a rotation of the magnetization around the $z$ axis in the laboratory frame. The effective eigenvalues of Hamiltonian (\ref{ESR-RTF}) are given by 
\begin{equation}
{\tilde E}_k=k \hbar\Omega_R,\quad -{N}\le k \le N,
\label{RabiE}
\end{equation}
which are equidistant $(\Delta {\tilde E}_k=\hbar\Omega_R)$, and again we consider that each spin interacts only with the field, individually. 

\section{Quantum treatment of spin-photon coupling}

Now we study the case where the interaction between spins and photons is treated quantum mechanically. We start by reviewing the energy diagram of model (\ref{JC-R}).

\subsection{Case of single spin $N=1$}

First, we consider the case of $N=1$. We adopt the basis $|n,\sigma\rangle$ where $n$ denotes the number of photons in the cavity, and $\sigma=-/+$ shows the ground/excited state as an eigenvalue of $\sigma^z$. The matrix of $\cal H$ is separated into $2\times 2$ blocks for each pair $\{|n,-\rangle,|n-1,+\rangle\}$, given by
\beq
\left(\begin{array}{cc} 
 {-\hbar\omega_0\over2}+n\hbar\omega & \hbar g\sqrt{n} \\
 \hbar g\sqrt{n} & {\hbar\omega_0\over2}+(n-1)\hbar\omega
\end{array}\right).
\eeq 
The eigenstates are given by
\beq
\begin{array}{rl}
E_{\pm}&=\hbar\omega(n-{1\over2})\pm \hbar{\delta\over2},\\ |\Psi_\pm\rangle&={1\over\sqrt{2}}
\left(\begin{array}{c} 
\sqrt{1\mp\Delta/\delta}\\
\pm\sqrt{1\pm\Delta/\delta}
\end{array}\right),
\end{array}
\label{eq9-10}
\eeq
where $\Delta=\omega-\omega_0$ and $\delta=\sqrt{\Delta^2+4ng^2}$. At resonance $\omega=\omega_0$, the above reduces to
\beq
\begin{array}{rl}
E_{\pm}&=(n-{1\over2})\hbar\omega_0\pm \hbar g\sqrt{n}, \\
|\Psi_{\pm}\rangle&={1\over\sqrt{2}}\left(\begin{array}{c} 
 1\\\pm 1\end{array}\right). 
\label{eq11}
\end{array}
\eeq
Note that all blocks have the same photon number + magnetization constant\cite{Agarwal_PRL84}, $C=n-1/2=(n-1)+1/2$.

\subsection{Case of a spin ensemble $N>1$}

For $N>1$, the working basis becomes $|n,\{\sigma_1\cdots\sigma_N \}\rangle$ where $\sigma_i=-/+$ shows the ground/excited state of $i$th spin, as an eigenvalue of $\sigma^z_i$. When photons are absorbed or emitted by the spin ensemble, the quantity $C=n+M$ is conserved~\cite{TCPR1968,Agarwal_PRL84}, with $M=m-N/2$ the ensemble magnetization, $m$ the number of excited spins and $n$ the number of remaining photons. The Hilbert space corresponding to all $m$ values is $_NC_0+\dots +_NC_N=2^N$ in length. Further division in independent sub-blocks can be done if one use a total spin representation \cite{Agarwal_PRL84}. The system Hamiltonian can be written as:
\begin{equation}
{\cal H}=\frac{\hbar\omega_0}{2}S^z+\hbar\omega b^{\dagger}b+\hbar g(bS^++b^\dagger S^-),
\label{HS}
\end{equation}
with $S^{z,+,-}=\sum_i^N\sigma_i^{z,+,-}$. The Hamiltonian commutes with the total spin $(S^x)^2+(S^y)^2+(S^z)^2$ and can be further separated into smaller blocks, classified by the total spin $S$. When all spins are in the ground state, the total spin is maximum, $S=N/2$. From here on we discuss the case $S=N/2$, since further photon excitations and emissions will selectively couple states within this subspace only. The subspace can be further limited if there are not enough photons to flip all the spins in the system: $m$ ranges from 0 to $min(N,n_{\rm max})$ with $n_{\rm max}=n+m$ the number of photons for $S_z=-N/2$. Using a basis  $\{|m\rangle, 0\le m\le min(N,n_{\rm max})\}$ and the relation ($S=N/2$):
\begin{equation}
\begin{array}{ll}
S^+|S,M\rangle & =\sqrt{S(S+1)-M(M+1)}|S,M+1\rangle= \\
& =\sqrt{(m+1)(N-m)}|S,M+1\rangle,
\end{array}
\end{equation}
the diagonal and off-diagonal terms of Eq. (\ref{HS}) are:
\begin{equation}\label{Hdiag}
\begin{array}{rl}
{\cal H}_{m,m}&=(m-N/2)\hbar\omega_0+(n_{\rm max}-m)\hbar\omega \\
{\cal H}_{m,m+1}&=\hbar g\sqrt{n_{\rm max}-m}\sqrt{(m+1)(N-m)}.
\end{array}
\end{equation}
At resonance, the diagonal term becomes ${\cal H}_{m,m}=(n_{\rm max}-N/2) \hbar\omega_0$, and is independent on $m$.

\section{Eigenvalues for an ensemble of spins}

For a total spin $S=N/2$, one expects $N+1$ eigenstates of which  $N-n_{max}$ are diagonal spin states and $n_{\rm max}+1$ are coupled spin-photon states (if $n_{\rm max}\ge N$, all $N+1$ states are coupled). 

\subsection{Analytical expressions for $n_{\rm max}=0-3$}
Analytical expressions for eigenvalues $E_m^{(n_{\rm max})}$ are listed below for few simple cases.

\subsubsection{\bf Case $n_{\rm max}=0$}.
In the vacuum-field where no photon exist when all spins are in the ground state, there is a unique state and its eigenvalue is given by
\beq
E^{(0)}_0=-{\hbar\omega_0 N\over 2}.
\label{e0}
\eeq 
\subsubsection{\bf Case $n_{\rm max}=1$}.
In this case there are two coupled states: $|n=1,m=N\rangle$ and  $|n=0,m=N-1\rangle$.
The Hamiltonian of this block is 
$$ \left(
\begin{array}{cc}
\hbar\omega_0\left(-{N\over2}\right)+\hbar\omega & g\sqrt{N} \\
g\sqrt{N} & \hbar\omega_0\left(-{N\over2}+1\right) 
\end{array} 
\right)
\begin{array}{l}
|n=1,m=0\rangle \\
|n=0,m=1\rangle \end{array}
$$
\beq
=\hbar\omega_0\left(-{N\over2}+1\right)
+ \left(
\begin{array}{cc} 
\hbar(\omega-\omega_0) & g\sqrt{N} \\
g\sqrt{N} & 0 \end{array} 
\right),
\eeq
and the eigenenergies are given by ($m=0,1$):
\beq
E_m^{(1)}/\hbar=\omega_0\left(-{N\over2}+1\right)+{\Delta\over2}+ \left( m-{1\over2}\right)\sqrt{\Delta^2+4Ng^2}
\label{e1}
\eeq

The VRS is given by
\beq
E_{1}^{(1)}-E_{0}^{(1)}=\hbar\sqrt{\Delta^2+4Ng^2}
\label{Rabi1}
\eeq
and represents the rate at which the spin system coherently exchanges one photon with the radiation field.

\subsubsection{\bf Case $n_{\rm max}=2$}. The spin-photon states span over three states: 
$|n=2,m=N\rangle$, $|n=1,m=N-1\rangle$, and $|n=0,m=N-2\rangle$. At resonance $\Delta=0$, the Hamiltonian is given by:
\beq \hbar\omega_0\left(2-{N\over2}\right)+\left(
\begin{array}{ccc}
0 & \hbar g \sqrt{2N} & 0 \\
\hbar g \sqrt{2N} & 0 & \hbar g \sqrt{2(N-1)}\\
0 & \hbar g \sqrt{2(N-1)}& 0 
\end{array}\right),
\eeq
and the eigenvalues are given by ($m=0,1,2$):
\beq
E^{(2)}_m/\hbar=\omega_0\left(-{N\over2}+2\right)+(m-1)g\sqrt{4N-2}
\label{e2}
\eeq
Here, the Rabi frequencies are degenerate
\beq
E^{(2)}_2-E^{(2)}_{1}=E^{(2)}_{1}-E^{(2)}_{0}=\hbar g\sqrt{4N-2}.
\label{Rabi2}
\eeq

\subsubsection{\bf Case $n_{\rm max}=3$}. Following a similar approach as for $n_{\rm max}=2$, the Hamiltonian at resonance is given by:
\beq 
\begin{array}{l}
\hbar\omega_0\left(3-{N\over2}\right)+\\
\left(
\begin{array}{cccc}
0 & \hbar g \sqrt{3N} & 0 & 0\\
\hbar g \sqrt{3N} & 0 & \hbar g \sqrt{4(N-1)} & 0\\
0 & \hbar g \sqrt{4(N-1)}& 0 &  \hbar g \sqrt{3(N-2)}\\
0 & 0 &  \hbar g \sqrt{3(N-2)} & 0
\end{array}\right),
\end{array}
\eeq
with eigenvalues:
\beq
\label{e3}
E_m^{(3)}/\hbar=\hbar\omega_0(-N/2+3)\pm g\sqrt{5(N-1)\pm\sqrt{(4N-5)^2+8N}},
\eeq
where $m=0,1,2$ and 3 counts the four possible values.

\subsection{General case $N, n_{\rm max}\ge 1$}
\begin{figure}
\includegraphics[width=\columnwidth]{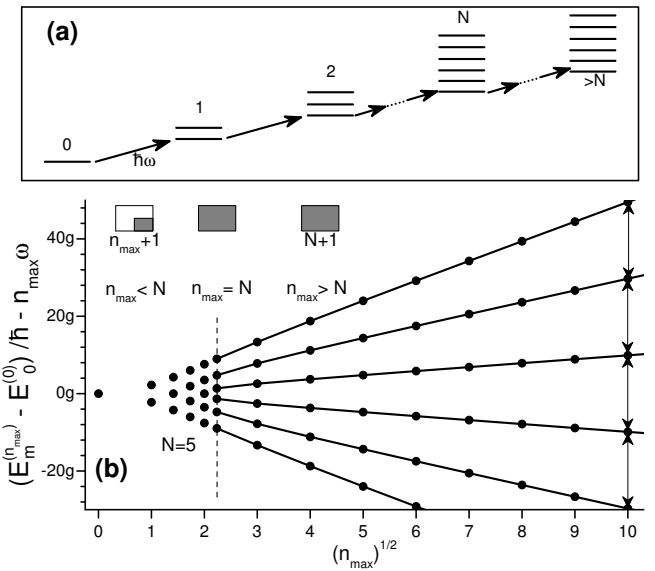}
\caption{{\rm (a)} For each additional photon, a new spin-photon state is generated, up to a number $N+1$. {\rm(b)} Spin-photon eigenstates for $N=5$, at resonance, obtained from diagonalization of blocks limited by $[min(n_{\rm max},N)+1]$, as sketched in the insert. For $n_{\rm max}\gg N$, all $N+1$ states become equidistant, leading to equal Rabi splittings (shown by vertical arrows).} 
\label{fig1}
\end{figure}

The size of the block representing the spin-photon states increases with $n_{\rm max}$. For each photon made available to the spin system, an additional spin will participate in the cooperative energy exchange and a new spin-photon state is generated, as indicated in Fig.~\ref{fig1}(a). This effect is exemplified in Fig.~\ref{fig1}, at resonance, for $N=5$ where up to six states are generated with increasing $n_{\rm max}$.  When $n_{\rm max}=5$, all spins are participating and the size of the matrix is bounded at $6\times 6$. Although no new states are generated for values of $n_{\rm max}>N$, the $N+1$ eigenvalues will adjust to indicate the ``oversaturation" with photons (as shown with connected dots in Fig.~\ref{fig1}). Analytically, this effect is shown by Eq.~(\ref{eq9-10}) for $N=1$ or by replacing $N$ with $n_{\rm max}$ in the off-diagonal terms leading to Eqs.~(\ref{e0}), (\ref{e1}), (\ref{e2}), and (\ref{e3}) and thus in the corresponding Rabi splittings. 

\begin{figure}
\includegraphics[width=\columnwidth]{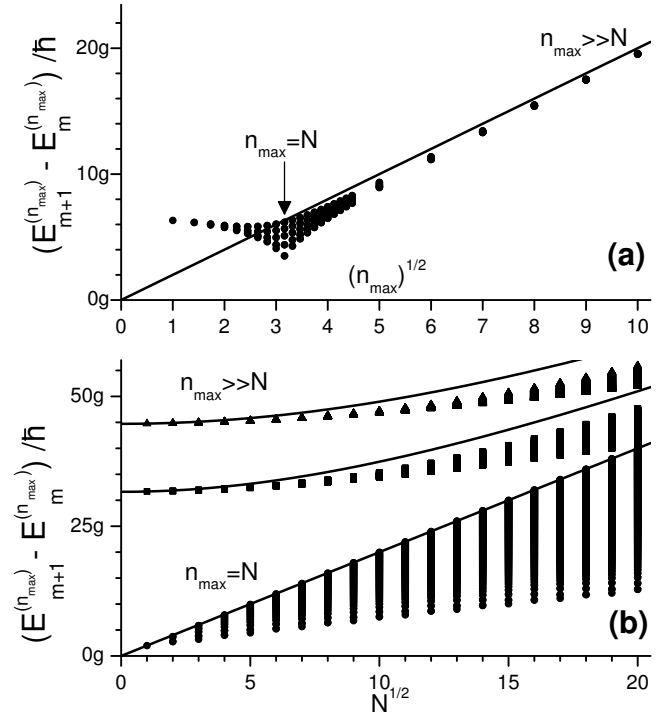}
\caption{Rabi splittings calculated at resonance, as difference between consecutive eigenvalues, for (a) $N=10$ and (b) $N$ up to 400 and $n_{\rm max}=N$ (dots), $N+250$ (squares), $N+500$ (triangles); the $N=1$ limit for Rabi splitting, $2\hbar g\sqrt{n_{\rm max}}$, is shown by continuous lines. At $n_{\rm max}=N$ the distribution width in Rabi splittings is maximal, and it decreases rapidly with increasing $n_{\rm max}$.} 
\label{fig2}
\end{figure}

For large values of $n_{\rm max}$, all Rabi splittings
\beq
\hbar\Omega_{\rm R}=E_{m+1}^{(n_{\rm max})}-E_m^{(n_{\rm max})}, 
\label{Rabi}
\eeq
are equal to $\approx 2\hbar g\sqrt{n_{\rm max}}$ (as shown by vertical arrows in the example of Fig.~\ref{fig1}). However, this transition to equidistance is not a smooth process. In the following, we show that the spread in Rabi frequencies becomes maximal at $n_{\rm max}=N$, and that a gradual transition towards an equidistant spectrum develops for $n_{\rm max}>N$. Difference between consecutive eigenvalues are shown in Fig.~\ref{fig2}. For $N=10$ [Fig.~\ref{fig2}(a)] one observes a spread of Rabi splittings over few units of $g$ for $n_{\rm max}=N=10$, followed by a collapse on a single-valued Rabi splitting $2\hbar g\sqrt{n_{\rm max}}$ [shown by continuous lines in Fig.~\ref{fig2}(a) and (b)] for $n_{\rm max}$ larger than several tens.

An exact diagonalization study for $N$ up to 400 and three $n_{\rm max}$ values provides support for a general view of the process [Fig.~\ref{fig2}(b)]. At $n_{\rm max}=N$ (black dots) deviations from the $2\hbar g\sqrt{n_{\rm max}}$ limit (black lines) are maximal and the range of spread increases proportionally with $\sqrt{N}$. For $n_{\rm max}=N+250$ and $N+500$ the distribution width of Rabi splittings is highly reduced. A convergence toward $2\hbar g\sqrt{n_{\rm max}}$ is observed especially at low values of $N$ (or very large $n_{\rm max}/N$ ratio), since in this limit one approaches the analytical case of $N=1$ [Eq.~\ref{eq11}]. 

This convergence corresponds to the energy diagram of the Rabi oscillation in Eq. (\ref{RabiE}) in the ESR model, where each atom interacts with the field individually. The fact that $\Omega_{R}\propto \sqrt{n_{\rm max}}$ in this limit, resides on the dependence $h_0\propto \sqrt{n_{\rm max}}$. If the number of photons is large enough, $b,b^{\dagger} \rightarrow \langle b\rangle, \langle b^{\dagger}\rangle \rightarrow \sqrt{n_{\rm max}} $ in Eq.(\ref{JC-R}).
Thus, $h_0$ in Eqs.~(\ref{ESR-RF}), (\ref{ESR-MX}) corresponds to $\sqrt{n_{\rm max}}$. 

\subsection{Photon transmission spectra: a quantum to classical transition}
 
To probe the $E_m^n$ spin-photon states one could use a low power beam and analyze the transmitted signal (another option, demonstrated experimentally in the last section, is to study the Fourier transform of the coherent emission of an excited spin-cavity system). A low power probe could be used to scan the frequency response of the cavity, after the introduction of $n_{\rm max}$ photons of frequency $\omega$ ($=\omega_0$ at resonance).

\begin{figure}
\includegraphics[width=\columnwidth]{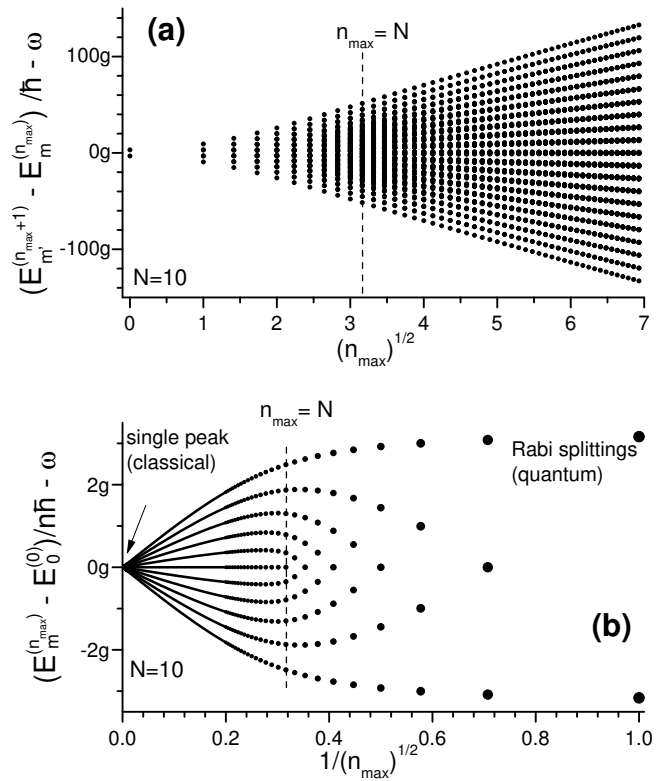}
\caption{{\rm (a)} Representation of all possible excitations from $n_{\rm max}$ to $n_{\rm max}+1$, calculated at resonance, for $N=10$. The corresponding energy splittings are closely packed in the transition region $n_{\rm max}\approx N$ and became equally spaced by $2\hbar g\sqrt{n_{\rm max}}$ for $n_{\rm max}\gg N$. {\rm (b)} Multiphoton, nonlinear resonances as a function of $1/\sqrt{n_{\rm max}}$. The linear dependence at high $n_{\rm max}$ indicates the emergence of an equally spaced Rabi spectra, and the convergence to a single peak indicates the gradual passage to a classical ESR, single-peaked, resonance condition.} 
\label{fig3}
\end{figure}

A variable frequency beam, probing after the introduction of $n_{\rm max}$ photons, will see the Rabi splittings given by Eq.~(\ref{Rabi}). Therefore, in such experiment one should be able to detect a transition in peak distribution from a large number of $\Omega_{R}$ values to a single-valued transmission peak. By design, such photon-driven transition cannot be observed at $N=1$ but it would require at least several non-interacting spins coupled to $n_{\rm max}\sim N$ photons.

The passage of a system from $n_{\rm max}$ to $n_{\rm max}+1$ implies an excitation from one group of eigenstates to the next one, as sketched in Fig.~\ref{fig1}(a). Due to the large number of eigenstates involved, the distribution of energy differences 
\beq
\hbar\Delta_{m'm}^{(n_{\rm max})}=E_{m'}^{(n_{\rm max}+1)}-E_m^{(n_{\rm max})}
\eeq
can be broad. As shown in Fig.~\ref{fig3}(a) for $N=10$, the distribution width is increasing as $\sqrt{n_{\rm max}}$. However, only in the vicinity of $n_{\rm max}=N$ are the values highly dispersed, as shown by a dense cloud of points at $n_{\rm max}\approx N$ in Fig.~\ref{fig3}(a). For large values of $n_{\rm max}$, and in full agreement with the study of Fig.~\ref{fig2}, the $\Delta_{m'm}^{(n_{\rm max})}$ values are equally spaced by a Rabi splitting $2\hbar g\sqrt{n_{\rm max}}$.

A different probing method would consist in using the $n_{\rm max}$ photons to actually probe the energy levels. An experimental demonstration for one superconducting qubit electrically coupled to $n_{\rm max}=0...5$ photons has been recently performed~\cite{Bishop_nphys09} (see also  Ref.~\onlinecite{Gripp_pra96} for a multi-atom experiment). In such a case, multi-photon transmission experiments can probe the energy difference between the $E_m^{n_{\rm max}}$ and $E_0^0$ [of Eq.~(\ref{e0})]
\beq
\hbar\omega=(E_m^{n_{\rm max}}-E_0^0)/n_{\rm max}.
\eeq
 Above the transition threshold $n_{\rm max}\approx N$, the photon frequency thus defined becomes gradually insensitive to the distribution in Rabi splittings  due to the $1/n_{\rm max}$ factor. A level diagram with Rabi splittings uniformly spaced by $2\hbar g\sqrt{n_{\rm max}}$ will generate transmission peaks spaced by $2\hbar g/\sqrt{n_{\rm max}}$. The linear dependence on $1/\sqrt{n_{\rm max}}$ for $n_{\rm max}\gg N$ is shown in Fig.~\ref{fig3}(b), calculated at resonance conditions for $N=10$. The gradual passage from the quantum case to the classical ESR condition (single peak at $\omega=\omega_0$) is visible with the increase of the electromagnetic intensity.

\section{Experimental study of strong coupling in spin systems}

We demonstrate the strong coupling between an ensemble of spins $s=1/2$ and photons using a sample of the well-known ESR standard material~\cite{Krzystek_JMR97}, dipheriyl-picri-hydrazyl (DPPH). For an optimized cavity coupling, sample positioning and size, we have been able to induce a sizeable Rabi splitting observed for a series of Zeeman spin splittings $\omega_0$.

\begin{figure}
\includegraphics[width=\columnwidth]{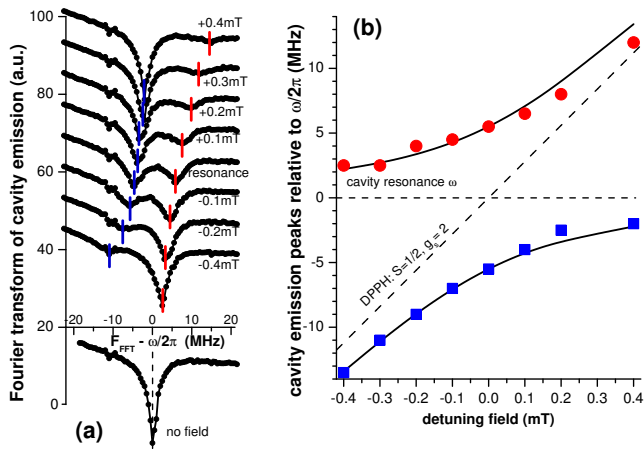}
\caption{(Color online) {\rm (a)} Fourier transform of cavity ringing, measured at room temperature for several magnetic field values around the resonance condition. Frequencies are relative to the sample-loaded cavity resonance, in absence of applied field. The observed two peaks are indicated by vertical marks. {\rm (b)} Peaks position as a function of detuning field $\mu_0\delta H_z$. The dashed lines indicate the classical level diagram, whereas the continuous line fit [Eq.~\ref{em}] shows a measured Rabi splitting of 10.9~MHz. } \label{fig4}
\end{figure}

A 1-$\mu$s-long microwave pulse pumps a large number of photons into a cylindrical cavity operated in mode TE$_{011}$. When the microwave is switched off, the cavity is coherently emitting photons corresponding to its own eigenmodes (phenomenon known as cavity ringing). This ringing is detected by a homemade heterodyne analyzer. The experiment is performed at room temperature. To ease the distinction between the photons of the pump pulse and cavity's own emitted photons after pumping, the pump is detuned by $50$~MHz from cavity's resonance $\omega/2\pi=9.624$~GHz. The Fourier transform of the coherent oscillations is shown in Fig.~\ref{fig4}(a), with the frequency axis shifted by $\omega/2\pi$ for clarity. The sample-loaded, no field ($\omega_0=0$) oscillation shows only the cavity signature, located at $\omega/2\pi$.   

For an applied magnetic field $\mu_0H_z=\hbar\omega/g_s\mu_B$, with $g_s=2$, the spin system is in resonance with the cavity ($\omega_0=\omega$). The eight traces of Fig.~\ref{fig4}(a) are for fields within $\pm 0.4$~mT of the resonance condition. The Fourier transform of the coupled spin-photon oscillations show peaks, indicated by vertical marks in Fig.~\ref{fig4}(a). The peaks location, relative to cavity's resonance $\omega/2\pi$, is shown in Fig.~\ref{fig4}(b), as a function of the field detuning $\mu_0\delta H_z=-\hbar\Delta/g_s\mu_B$.

In a classical ESR experiment, for instance using the DPPH as a field calibration standard, one expects energy levels that follow the dashed lines in Fig.~\ref{fig4}(b). In particular, the cavity peak is visible and changes abruptly in size but not location, when the resonance condition is met (at the intersection of the dashed lines). 

In our experimental conditions, due to the spin-photon strong coupling, one observes two peaks separated by the characteristic Rabi splitting. The size of the splitting can be attributed to the interaction between a group of $N$ spins and a single photon, although the cavity contains a large number of $n_{\rm max}$ photons. A possible explanation resides on the existence, in standard ESR experiments, of the so-called ``spin packets" characterized by a coherence time $T_2$ and grouping an average number of spins $N$. The $T_2$ and cavity decay times can be estimated from the peak widths of $\approx 6$ MHz and 2.7~MHz at $\omega=\omega_0$ (resonance) and $\omega_0=0$ (no field) conditions, respectively: $T_2= 170$~ns and $\tau_{cav}= 370$~ns.

Consequently, the cavity photon depletion during emission, marks the transition between the $E_{0,1}^{(1)}$ and $E_0^{(0)}$ levels of Eqs.~(\ref{e1}),and (\ref{e0}). Cavity ringing shows coherent oscillations with the frequencies $\omega_{e0,1}/2\pi$, given by:
\beq
\omega_{e0,1}-\omega=-{\Delta\over2}\pm{1\over2}\sqrt{\Delta^2+\Omega_R^2},
\label{em}
\eeq
on which the continuous lines of Fig.~\ref{fig4}(b) are based. The fit procedure leads to a Rabi splitting of $\Omega_R/2\pi=10.9$~MHz. We note that a detailed knowledge of the $g$ and $N$ parameters would require an on-chip type of experiment, at low temperatures, to maximize the "spin packet" size and to ensure a precise knowledge of spin position \cite{Groll_JAP09}. In a single photon picture, and knowing the cavity volume ($\sim 50$~cm$^3$), we estimate the number of spins contributing to $\Omega_R$ to be on the order of $10^{20}$.  Our data demonstrate the spin-photon strong coupling and also provide the mainframe to study experimentally the transition shown in Fig.~\ref{fig3}(b).  The role of other factors in the spread of Rabi splittings, such as: dipolar or hyperfine interactions, local anisotropic crystal fields, and the size of the sample vs. size of a cavity (mode) can be studied as well. At the same time, the possibility to entangle spins and photons inside a cavity comes in strong support for the implementation of quantum computing algorithms by using magnetic spins and on-chip quantum electro-dynamics methods.

In conclusion, we present a study of cooperative spin-photon interaction leading to a transition between a quantum-type Rabi spectra to a classical ESR spectra. The transition requires a number of spins $N>1$, and the possibility to gradually increase the number of photons. When the later is close to $N$, a dense Rabi spectra is numerically observed. For a large number of photons, the spectra gradually become equidistant and the resonance peaks converge toward the classical single-peaked, ESR resonance. The theoretical model is complimented by an experimental demonstration of the spin-photon strong coupling regime, which uses the B-field component of an electromagnetic mode in a cylindrical cavity.\\
{\it Note added}. Recently, two similar experimental results have appeared \cite{Schuster,Kubo}.

\section*{Acknowledgments}\label{sec:ACK}
This work was supported by the NSF Cooperative Agreement Grant No. DMR-0654118 and No. NHMFL-UCGP 5059, NSF grant No. DMR-0645408, the Alfred P. Sloan Foundation, ``Physics of new quantum phases in superclean materials'' (Grant No. 17071011), and also the Next Generation Super Computer Project, Nanoscience Program of 
MEXT. Numerical calculations were done on the supercomputer of ISSP.

\end{document}